# GOING BEYOND BELL'S THEOREM

Daniel M. Greenberger[1], Michael A. Horne[2], and Anton Zeilinger[3].
[1]City College of the City University of New York, New York, New York
[2]Stonehill College. North Easton, Massachusetts
[3]Atominstitut der Oesterreichischen Universitaeten, Wien, Austria

ABSTRACT. Bell's Theorem proved that one cannot in general reproduce the results of quantum theory with a classical, deterministic local model. However, Einstein originally considered the case where one could define an "element of reality", namely for the much simpler case where one could predict with certainty a definite outcome for an experiment. For this simple case, Bell's Theorem says nothing. But by using a slightly more complicated model than Bell, one can show that even in this simple case where one can make definite predictions, one still cannot generally introduce deterministic, local models to explain the results.

In 1935 Einstein, Podolsky, and Rosen (1) wrote their classic paper (EPR) which pointed directly to the Achilles' Heel of quantum theory. They pointed out that if quantum theory were true, it would have to defy common sense in a manner which was very distasteful to a classically oriented mind. Bohr's answer (2) was not a refutation of their logic, but rather an affirmation of the fact that quantum theory does just that. The subsequent history of the subject, which has vindicated Bohr, is not to be taken as a refutation of EPR, but rather as a confirmation of just exactly how counter-intuitive a theory quantum theory is. An indication of how expertly they zeroed in on the most troubling aspect of the subject is the fact that in 1985 alone, 50 years after their paper was written, there were still 48 journal citations of their original article.

They were interested in the completeness of the theory, and they defined a complete theory as one in which "Every element of the Physical Reality must have a counterpart in the physical theory". As to the phrase "Physical Reality" that occurs here, they made no claim to be able to define it in general. Rather, they gave what they thought should be one minimal requirement that an element of physical reality should exhibit. It is this requirement, which seems so necessary and obvious, that quantum theory violates. They proposed that "if, without in any way disturbing a system, we can predict with certainty (i.e., a probability equal to unity) the value of a physical quantity, then there exists an element of physical reality corresponding to this physical quantity."

They gave an example, but most subsequent discussion has used a different example given by Bohm (3). Consider a spin-0 system which decays into two spin ½ particles. The wave function will be

$$|\psi\rangle = \left(|\uparrow\downarrow\rangle - |\downarrow\uparrow\rangle\right)/\sqrt{2}.$$

so that if particle 1 comes off with spin up (↑), particle 2 will have spin down (↓), and vice-versa. The two particles will come off in opposite directions to conserve momentum. If one measures the spin of particle 1, far from the decay point, and finds spin up, say, then one knows with certainty that particle 2, which is far away, has spin down. According to the EPR argument, since one has in no way disturbed particle 2, then this feature, spin down, must be an element of physical reality. Therefore having spin down is a property of the particle itself, and cannot have been produced by any measurement we made on particle 1. It must have come away from the point of interaction, the decay point, with spin down.

Quantum mechanics denies this simple point. It says that the spin of particle 2 is indeterminate until the spin of particle 1 is measured, as until then it was in a superposition of states up and down, and one could in principle have interference between the possibilities. This was the crux of the dispute between Einstein and Bohr, but it was thought until 1965 that the difference between the two points of view had no experimental consequences. Only then did Bell prove his famous theorem (4) that in fact the assumption of the reality of the spin places severe restrictions of the possible correlations that can exist between the particles, if one makes spin measurements in arbitrary directions. Many experiments done since then have confirmed the results of quantum theory.

But it is interesting that Bell's results say nothing in the special case covered directly by the EPR argument, namely the case where a measurement on one particle allows one to predict what happens to the other particle with 100% certainty. This is the case where one measures the spin on one particle, and then measures the other either in the same or opposite direction. Not only does this case yield certainty in its measurement, but in fact one can arrive at a classical model of the system which gives the same result. It is only in the general case of an arbitrary angle between the particles, where one does not have certain knowledge, that quantum theory yields results that contradict the classical ones.

Specifically, with the wave function above, if one measures the spin of particle 1 in some direction **n**, while one measures the spin of particle 2 in a different direction **l**, then the expectation of the correlation between the two particles quantum mechanically will be

$$E(n \cdot l) = \langle\psi|(\sigma \cdot n)(\sigma \cdot l)|\psi\rangle = -\cos(n \cdot l),$$

and in the case where the particles are moving along the ± z direction, while **n** and **l** are in the x-y plane at angles α and β, this becomes cos (α − β). The cases where a definite prediction is possible are given by those mentioned above, where the measurement directions differ by 0° or 180°. We call this case the "super-classical" case, where an element of reality exists by virtue of perfect predictability, according to the EPR criterion.

In constructing a model for correlations in the case of a deterministic and local theory, Bell assumed that the spin of the particles were determined at the point they separated, according to EPR. Since the measurement of the spin in a given direction can only give two possible values, he assigned a value ± 1 to the result. Thus he gave as the result of a measurement of the spin of both particles, one along **n** and the other along **l**, the value

$$A_\lambda(n) \cdot B_\lambda(l),$$

where both A and B could have the values only ± l, which in a particular case were determined by some internal, hidden, variable λ. The only limitation on the product was that, as stated above, if **l = ± n**, then one had

$$A_\lambda(n) \cdot B_\lambda(n) = -1, \qquad A_\lambda(n) \cdot B_\lambda(-n) = +1.$$

Finally, the expectation value of the measurement represented the weighted sum over all possibilities λ,

$$E(n, l) = \int d\lambda\, \rho(\lambda)\, A_\lambda(n)\, B_\lambda(l).$$

Because of the factorable nature of the probabilities, he was able to derive the inequality

$$|E(n, l) - E(n, k)| \leq 1 + E(l, k).$$

This inequality is broken by the quantum mechanical result for most angles. But unfortunately, it gives no information at all when **l = ± n**.

In the super-classical case, where one can make a definite prediction, but where the Bell inequality above gives no information, one can make a simple deterministic model to explain the result, but of course, the Bell inequality shows that this is not possible for general angles. However it leads naturally to the question "Can one always find a classical model for the superclassical case?" While that is the trivial case from the point of view of Bell's inequality, it is the most interesting from the point of view of reality. In other words, Bell's theorem answers the question of whether one can make a classical, local, deterministic model to duplicate the results of quantum theory in general, and the answer is no. But it does not address the question of whether one can make such a model in the special case when one can make definite predictions, the EPR case.

The answer to this question is also no. But in order to answer it one needs a more complicated model than the two-body decay above. We have constructed a simple generalization of the Bohm model, which provides a greater restriction on the possibilities for the various particles. In fact, the restriction is so great that even in the super-classical case, one cannot make a deterministic, local model, and one does not even need inequalities. For any given value of the hidden variable λ one can show that it is impossible to construct such a model.

Our general attitude is that we assume that quantum mechanics gives correct answers, and the question is whether a classical model can reproduce these answers. The example we have chosen is of a particle of spin 1, initially the state

m=0 in a magnetic field along the z axis. The particle then decays into two particles, each of spin 1, one along the +z axis and on along the –z axis. Subsequently, each of these two particles decays into two spin ½ particles, the first two also moving along the +z axis, and the other two along the –z axis. We introduce this restriction so as not to introduce any spin-orbit type considerations into the problem.

In this problem the initial particle decays into four spin ½ and its wave function can be expressed as

$$|1, 0\rangle = (|\uparrow\uparrow\downarrow\downarrow\rangle - |\downarrow\downarrow\uparrow\uparrow\rangle) / \sqrt{2}.$$

The quantum mechanical expectation value for the spins in four given directions is

$$E(n_1, n_2, n_3, n_4) = -\cos(\alpha + \beta - \gamma - \delta),$$

where each of the directions $n_i$ is assumed to be in the x-y plane at angles $\alpha, \beta, \gamma, \delta$, respectively. Note that if any two of the angles are fixed, the other two obey the same law as the two body decay before, so that they will obey the Bell inequality. But the important point for us is that if

$$\alpha + \beta + \gamma + \delta = 0, \pi,$$

then the cos term will equal ±1, and so if we measure three of the angles, we can predict the fourth with 100% certainty. This is exactly the super-classical EPR case!

Thus we have again a general case where for most angles, we can make no specific classical prediction, however there is a range of parameters, given by the equation above, for which we can indeed make a definite prediction. The next question is whether we can find a classical, deterministic, local model for it. Since as before, if we measure the spin in any specific direction, we can get only two answers, we can use the same type of parametrization as before. We then get

$$A_\lambda(\alpha) \, B_\lambda(\beta) \, C_\lambda(\gamma) \, D_\lambda(\delta)$$

as our measure of the four particles landing at different angles. (The hidden variable λ can stand for any configuration of hidden parameters. But in this case they are all determined back at the original decay. The subsequent decay hidden variables will be determined by the original hidden variables of the first decay.) The condition corresponding to the super-classical case is

$$A_\lambda(\alpha) \, B_\lambda(\beta) \, C_\lambda(\gamma) \, D_\lambda(\delta) = \pm 1,$$
$$\text{for } \alpha + \beta + \gamma + \delta = 0, \pi \quad .$$

But it turns out that there is <u>no</u> way to satisfy this condition. It is too restrictive, because we can continuously vary two of the parameters while keeping the other two constant. This leads to the conclusion that A=B=C=D=constant. But this is impossible, since the product sometimes equals +1 and sometimes equals -1. This is true for any value of λ so that there is no need to integrate over it. Thus we reach the general conclusion that not only is there no way to form a classical, deterministic, local theory that reproduces quantum theory in general, but that even in the simpler case that one can make definite predictions in the EPR sense, it is impossible to do so with such a model. However one must go beyond the Bell theorem in order to prove this. A further conclusion is that with the appropriate 4-particle (or even 3-particle) system, all one must do is prove that quantum theory holds experimentally, and then we know that it cannot be classically duplicated, so that it will be much easier to disprove the classical type of loop-holes that are constantly being sought to explain the results of 2-particle experiments which verify quantum theory.

ACKNOWLEDGEMENT. We would like to thank the National Science Foundation and the Humboldt Stiftung (DMG) for providing partial support for this work.